\documentclass[twoside,11pt]{article}
\usepackage{jmlr2e}
\ShortHeadings{Dual Lasso Selector}{Niharika Gauraha}
\firstpageno{1}

\usepackage{fancyhdr,amsmath,amssymb,amsfonts}%
\usepackage{caption}
\usepackage[ruled]{algorithm2e}
\usepackage{booktabs,tabu}

\usepackage{subfiles}

\begin{document}

\title{Dual Lasso Selector}
\author{\name Niharika Gauraha  \email niharika.gauraha@gmail.com \\
       \addr Systems Science and Informatics Unit\\
       Indian Statistical Institute\\
       8th Mile, Mysore Road Bangalore, India\\
       }

\editor{}

\maketitle

\begin{abstract}
We consider the problem of model selection and estimation in sparse high dimensional linear regression models with strongly correlated variables. First, we study the theoretical properties of the dual Lasso solution, and we show that joint consideration of the Lasso primal and its dual solutions are useful for selecting correlated active variables. Second, we argue that correlations among active predictors are not problematic, and we derive a new weaker condition on the design matrix, called Pseudo Irrepresentable Condition (PIC). Third, we present a new variable selection procedure, Dual Lasso Selector, and we prove that the PIC is a necessary and sufficient condition for consistent variable selection for the proposed method. Finally, by combining the dual Lasso selector further with the Ridge estimation even better prediction performance is achieved. We call the combination (DLSelect+Ridge), it can be viewed as a new combined approach for inference in high-dimensional regression models with correlated variables. We illustrate DLSelect+Ridge method and compare it with popular existing methods in terms of variable selection, prediction accuracy, estimation accuracy and computation speed by considering various simulated and real data examples. \end{abstract}

\begin{keywords}
Dual Lasso Selector, Correlated Variable Selection, High-dimensional Statistics, Lasso, Lasso Dual, Ridge Regression
\end{keywords}

\section{Introduction and Motivation}
The use of microarray technologies have become popular to monitor genome-wide expression changes in health and disease. Typically, a microarray data set is high dimensional in the sense, it usually has tens of thousands of gene expression profile(variables) but only tens or hundreds of subjects(observations). In microarray analysis, a group of genes sharing the same biological pathway tend to have highly correlated expression levels \cite{Segal} and the goal is to identify all(rather than a few) of them if they are related to the underlying biological process. This is one example where the need to select groups of correlated variables arises. In many applications it is required to identify all relevant correlated variables. 
In this paper, we consider the problem of model selection and estimation in sparse high dimensional linear regression models with strongly correlated variables.

We start with the standard linear regression model as
\begin{align} \label{eq:lr}
	\textbf{Y} &=  \textbf{X} \beta+\epsilon,
\end{align}
with response vector $\textbf{Y}_{n \times 1}$, design matrix $\textbf{X}_{ n \times p}$, true underlying coefficient vector $\beta_{p \times 1}$ and error vector $\epsilon_{n\times 1} \sim N_n(0,I_n)$. In particular, we consider the case of sparse high dimensional linear model $(p \gg n)$ with strong empirical correlation among few variables. The Lasso is a widely used regularized regression method to find sparse solutions, the lasso estimator is defined as
\begin{align}
	 \hat{\beta}_{Lasso} = \arg\min_{\beta \in \mathbb{R}^p} \left\lbrace \frac{1}{2}\|{ \textbf{Y} -\textbf{X} \beta}\|_2^2 + \lambda \|\beta\|_1 \right\rbrace, \label{eq:lasso}
\end{align}
where $\lambda \geq 0$ is the regularization parameter that controls the amount of regularization. It is known that the Lasso tends to select a single variable from a group of strongly correlated variables even if many or all of these variables are important.

In presence of correlated predictors, the concept of clustering or grouping correlated predictors and then pursuing group-wise model fitting was proposed, see \cite{Buhlmann2} and \cite{Niharika}. When the dimension is very high or in case of overlapping clusters, finding an appropriate group structure remains as difficult as the original problem. We note that clustering followed by model fitting is computationally expensive, not reliable and do not scale for large, high-dimensional data sets, so we do not consider it further in this paper. An alternatively approach is simultaneous clustering and model fitting that involves combination of two different penalties. For example, Elastic Net (\cite{Hui}) is a combination of two regularization techniques, the $\ell_2$ regularization provides grouping effects and $\ell_1$  regularization produces sparse models. Therefore, the eNet selects or drops highly correlated variables together that depends on the amount of $\ell_1$ and $\ell_2$ regularization. 

The influence of correlations on Lasso prediction has been studied in \cite{Hebiri} and \cite{vandegeer}, and it is shown that Lasso prediction works well in presence of any degree of correlations with an appropriate amount of regularization. However, studies show that correlations are problematic for parameter estimation and variable selection. It has been proven that the design matrix must satisfy the following two conditions for the Lasso to perform exact variable selection: irrepresentability(IC) condition(\cite{Zhao}) and beta-min condition(\cite{Buhlmann1}). Having highly correlated variables implies that the design matrix violates the IC, and the Lasso solution is not stable. When active covariates are highly correlated the Lasso solution is not unique and Lasso randomly selects one variable from correlated group.  However, even in case of highly correlated variables the corresponding dual Lasso solution is always unique. The dual of the Lasso problem (\ref{eq:lasso}), as shown in \cite{Jie} is given by 
\begin{align} \label{eq:dualLasso}
	\sup_{\theta} \; & g(\theta) =  \frac{1}{2} \|\textbf{Y} \|_2^2 - \| \theta - \textbf{Y} \|_2^2 \nonumber \\
	\text{subject to } &  |X_j^T \theta| \leq \lambda \; \text{ for all } j \in \{ 1,...,p \} 
\end{align}
The intuition drawn from the articles \cite{Osborne} and \cite{Jie} further motivates us to consider the Lasso optimal and its dual optimal solution together, that yields in selecting correlated active predictors. 

Exploiting the fact about uniqueness of the dual Lasso solution, we propose a new variable selection procedure, the Dual Lasso Selector (DLS). For a given $\lambda$ and a Lasso estimator $\hat{\beta}_{Lasso}$, we can compute the corresponding Dual Lasso solution using KKT conditions.  Basically, the DLS active set corresponds to the predictors that satisfies dual Lasso feasible boundary conditions (we discuss it in details in later section). We argue that correlations among active predictors are not problematic, and we define a new weaker condition on the design matrix that allows for correlation among active predictors, called Pseudo Irrepresentable Condition (PIC). We prove that the Pseudo Irrepresentable Condition is a necessary and sufficient condition for the proposed dual Lasso selector to select the true active set (under assumption of beta-min condition) with high probability. Moreover, we use the $\ell_2$ penalty (the Ridge regression, \cite{Ridge}) which is known to perform best in case of correlated variables, to estimate the coefficients of the predictors selected by the dual Lasso selector. We call the combination of the two, the DLSelect+Ridge. Though, DLSelect+Ridge resembles the "Ridge post Lasso" but it is conceptually different and behaves differently than the Lasso followed by the Ridge, especially in the presence of highly correlated variables. For example DLSelect+Ridge looks like Elastic-net, since both are combination of $\ell_1$ and $\ell_2$ penalties but Elastic-net is a combination of the Ridge Regression followed by the Lasso. In addition, Enet needs to cross-validate on a two-dimensional surface $O(k^2)$ to select its the optimal regularization parameters, whereas DLSelect+Ridge needs to cross validated twice on one-dimensional surface $O(k)$, where k is the length of the search space for a regularization parameter. 

Our contribution is summarized as follows:
\begin{enumerate}

\item We briefly review the state-of-the-art methods of simultaneous clustering and model fitting using combination of penalties such as Elastic-net, OSCAR and Fused Lasso etc.

\item We study the theoretical properties of the Lasso and its dual optimal solution together and we show that selection of active correlated variables is related to the dual feasible boundary conditions. 

\item By further exploiting the uniqueness property of the dual Lasso solution, we develop a variable selection algorithm to efficiently select the true active predictors (including correlated active predictors). we call this selection technique as the  \textit{Dual Lasso Selector}.

\item We derive the Pseudo Irrepresentable Conditions (PIC) for the design matrix that allow for the correlation between active covariates, and we show that under assumption of PIC the dual Lasso selector is variable selection consistent.

\item We propose a new combined approach, the DLSelect+Ridge: Dual Lasso selecting predictors and the Ridge estimating their coefficients.

\item We study the theoretical properties of the combination  DLSelect+Ridge.


\item We implement the DLSelect+Ridge method and empirically compare it with existing methods like Lasso and Enet etc. in terms of variable selection consistency, prediction accuracy, estimation accuracy and time complexity (using various simulations and real data examples).
\end{enumerate}

 

We have organized the rest of the article in the following manner. We start with background in section 2. In section 3, we present Dual Lasso Selector. We define PIC and discuss variable selection consistency under this assumption on the design matrix, in section 4. Section 5 is concerned with illustration of the proposed method on real and simulated data sets. Section 6 gives computational details. We shall provide some concluding remarks in section 7. 

\section{Notations and Background}
In this section, we state notations and assumptions, used throughout the paper.

We consider usual sparse high-dim linear regression model as given in \ref{eq:lr} with $p \gg n$. For the design matrix $\textbf{X} \in \mathcal{R}^{n \times p}$, we represent rows by $x_i^T \in \mathbb{R}^p, \ i= 1,...,n$, and columns by $X_j^T \in \mathbb{R}^n, \ i= 1,...,p$. We assume that the design matrix $\textbf{X}_{n\times p}$ is fixed, the data is centred and the predictors are standardized, so that $\sum_{i=1}^{n} \textbf{Y}_i = 0$, $\sum_{i=1}^{n} ({X}_{j}){i} = 0$ and $\frac{1}{n} \textbf{X}^T_{j}\textbf{X}_{j} = 1$ for all $j=1,...,p$. We denote by
 \begin{align}
 	S = \{j \in \{ 1,...,p \} : \beta_{j} \neq 0\},
 \end{align}
the true active set and cardinality of the set $s = |S|$, is called sparsity index. We assume that the true coefficient vector $\beta$ is sparse, that is $s \ll p$. We denote $\textbf{X}_S$ as the restriction of $\textbf{X}$ to columns in $S$, and $\beta_S$ is the vector $\beta$ restricted to the support $S$, with 0 outside the support $S$. Without loss of generality we can assume that the first $s$ variables are the active variables, and we partition the covariance matrix, $C = \frac{1}{n} \textbf{X}^{T}\textbf{X}$, for the active and the redundant variables as follows.
\begin{align} \label{eq:sigma_par}
	C = \left[ \begin{array}{cc}
	C_{11}  & C_{12}\\
	C_{21} & C_{22}	
	\end{array} \right]
\end{align}
Similarly the coefficient vector $\beta$ can be partitioned as $ \left[ \begin{array}{c}
	\beta_1 \\
	\beta_2	
	\end{array} \right]. $\\
The $\ell_1$-norm and $\ell_2$-norm (square) are defined as
\begin{align}
	\|\beta\|_1 &= \textstyle \sum_{j=1}^p |\beta_j|  \label{eq:l1} \\
	\|\beta\|_2^2 &= \textstyle \sum_{j=1}^p \beta_j^2.
\end{align}
Throughout the paper, we use the notation $\lambda_1 > 0$ for $\ell_1$ penalty and $\lambda_2  > 0$ for other penalty functions. 
For a vector $a \in \mathbb{R}^p$, we denote its sign vector as
\begin{equation}
\begin{aligned}
 \mathbb{S}(a) =  \left\lbrace \begin{array}{ll}
	1 & \text{ if } a  > 0  \\
	-1 &  \text{ if } a  < 0 \\
	0  & \text{ if } a  = 0  \\
\end{array}	 \right.
\end{aligned}
\end{equation}
We denote sub-gradient of $\ell_1$-norm evaluated at $\beta \in \mathbb{R}^p$, as $ \tau \in \partial \|\beta\|_1$, where $\tau$ satisfies the following. 
\begin{align}
	\tau_i = \left\lbrace \begin{array}{ll}
	1 & \text{ if } \beta_i   > 0   \\
	\left[-1,1 \right] & \text{ if }  \beta_i = 0\\
	-1 &  \text{ if } \beta_i  < 0 \\
\end{array}	 \right.
\end{align}

 
 \section{Review of Relevant Work}
Given the huge literature on the use of Lasso-type penalties for variable selection, we provide only a brief overview here, with focus on previous approaches which are closely related to our work. In particular, we briefly review the Lasso, the Ridge and the state-of-the-art in simultaneous clustering and model fitting using combination of penalties for high-dimensional sparse linear models. In general, we define a penalized least squares method as follows.

\begin{align}
	 \min_{\beta \in \mathbb{R}^p} \left\lbrace \frac{1}{2}\|{ \textbf{Y} -\textbf{X} \beta}\|_2^2 + \mathcal{P}_{method}(\beta, .) \right\rbrace \label{eq:pen_ls}
\end{align}
 where the penalty terms $ \mathcal{P}_{method}(\beta, .)$ can be different for different methods depending on the type and number of penalties used. In the following we define various penalized least squares estimators in terms of penalties used by them, and we also mention their computational complexity, variable selection consistency and grouping effects of selecting and dropping highly correlated predictors together.
\begin{enumerate}
\item Lasso: The Lasso method was proposed by \cite{Tibshirani} and the lasso penalty is defined as  
\begin{align}
	 \mathcal{P}_{Lasso}(\beta, \lambda_1) = \lambda_1 \|\beta\|_1  \label{eq:lasso_pen}
\end{align}

It uses the single $\ell_1$ penalty, and due to nature of the $\ell_1$ penalty it simultaneously performs variable selection and estimation. The whole regularization path can be computed efficiently with the computational effort of a single OLS fit, by some modification of the LARS algorithm, see \cite{efron2004}. It does not provide grouping effect, in fact the Lasso tends to select a single predictor from a group of highly correlated predictors. 

\item Ridge Regression (RR):  The ridge method was proposed by \cite{Ridge} and the ridge penalty is defined as follows.
\begin{align}
	  \mathcal{P}_{Ridge}(\beta, \lambda_2) =  \lambda_2 \|\beta\|_2  \label{eq:ridge}
\end{align}

It uses the single $L_2$ penalty, and it always has a unique solution for a fixed regularization parameter $\lambda_2$. Though it is known to correctly detect the variable signs with reduced mean square error with correlated variables, but does not provide variable selection. It provides grouping effect with the highly correlated variables, and the computational complexity of the ridge is same as the computational effort of a single OLS fit.

\item Elastic-net (Enet): The Enet method was introduced by \cite{Hui}. The Enet penalty is a combination of $\ell_1$ and $L_2$ penalties and it is defined as follows.
\begin{align}
	  \mathcal{P}_{Enet}(\beta, \lambda_1, \lambda_2) =  \lambda_1 \|\beta\|_1 + \lambda_2 \|\beta\|_2  \label{eq:enet}
\end{align}
Enet addresses both the limitations of the Lasso, that is it can select correlated predictors as well as it can handle the $s>n$ case. It provides grouping effect, but requires to search in two-dimensional space for choosing optimal values of its regularization parameters. Hence its effective time complexity depends on the length of the search space for the regularization parameters $\lambda_1$ and $\lambda_2$.

\item Correlation Based Penalty (CP): The correlation based penalized least squares method was proposed by \cite{Tutz2009}, 
which uses the following correlation-based penalty term
\begin{align*}
	\mathcal{P}_{CP}(\beta, \lambda_2) = \lambda_2 \sum_{i=1}^{p-1} \sum_{j > i}  \left\lbrace \frac{(\beta_i - \beta_j)^2 }{1 - \rho_{ij}}  +\frac{(\beta_i + \beta_j)^2 }{1 + \rho_{ij}} \right\rbrace
\end{align*}
It uses the single $CP$ penalty norm, and it always has a unique solution for a fixed regularization parameter and the grouping effect strongly depends on the convexity of the penalty term. It does not provide variable selection. However, a boosted version of the penalized estimator allows to select variables. But the major drawback is that it is not scalable for large high dimensional problems.

\item Fused Lasso: The Fused Lasso method was given by \cite{Fused}, and the Fused Lasso penalty is defined as
\begin{align}
	 \mathcal{P}_{Fused}(\beta, \lambda_1, \lambda_2) =  \lambda_1 \|\beta\|_1 + \lambda_2 \sum_{j=2}^{p} | \beta_j - \beta_{j-1} |  \label{eq:fused}
\end{align}
The first constraint encourages sparsity in the coefficients and the second constraint encourages sparsity in their differences.
The major drawback of this method is that it requires the covariates to be in some order. 
It does not perform automated variable clustering to unordered features.

\item OSCAR: The OSCAR was invented by \cite{Oscar}, and the OSCAR penalty is given as follows.
\begin{align*}
	  \mathcal{P}_{Oscar}(\beta, \lambda_1, \lambda_2)= \lambda_1 \|\beta\|_1 + \lambda_2 \sum_{i<j} max\{| \beta_i| , | \beta_{j} | \}  
\end{align*}
with $|\beta_1| \leq ... \leq |\beta_p|$.
The first constraint is to encourage sparsity in the coefficients and the second constraint encourages equi-sparsity in $|\beta|$. 
The time complexity limits its scalability on ultra high-dimensional problems, moreover it requires two-dimensional grid search over the two parameters $(\lambda_1 , \lambda_2)$ 

\item L1CP : The L1CP penalty term is given by as follows, see     \cite{L1CP}.
\begin{align}
	\mathcal{P}_{L1CP}(\beta, \lambda_1, \lambda_2) = \lambda_1 \| \beta \|_1 + \lambda_2 \sum_{i=1}^{p-1} \sum_{j > i}  \left\lbrace \frac{(\beta_i - \beta_j)^2 }{1 - \rho_{ij}} + \frac{(\beta_i + \beta_j)^2 }{1 + \rho_{ij}} \right\rbrace . \label{eq:l1cp}
\end{align}
It performs variable selection with grouping effect and estimation together but is not scalable to the large scale problems due to expensive computation time, and it also requires two-dimensional grid search over the two parameters $(\lambda_1 , \lambda_2)$.

\item Clustered Lasso: 
The Clustered Lasso penalty is defined as 
\begin{align*}
	 \mathcal{P}_{CL}(\beta, \lambda_1, \lambda_2) = \lambda_1 \|\beta\|_1 + \lambda_2 \sum_{i<j} | \beta_i - \beta_{j} | 
\end{align*}
The first constraint encourages sparsity in the coefficients and the second constraint encourages equi-sparsity in $|\beta|$, It is similar as the Fused lasso but does not require ordering of variables, see \cite{she2010}. It provides grouping effect, but it requires two-dimensional grid search over the two parameters $(\lambda_1 , \lambda_2)$. It is computationally expensive since it has to check equi-sparsity pattern for each pair of variables.


\end{enumerate}

In the following table we summarize the properties discussed above for various regularization methods.

\begin{small}
\begin{table}[h!] 
		\begin{tabular}{ p{35mm} p{20mm} p{20mm} p{20mm} p{20mm} p{20mm} }
		\toprule 
		 Properties\;\;\;/\ \;\;Methods & Clustering/ Ordering Required & Variable Selection & Grouping Effect & Scalability & Grid Search\\
		 \toprule
		 Lasso &  No & Yes & No & Yes & 1D \\
		 Ridge &  No & No & Yes & Yes & 1D \\
		 PC &  No & No & Yes & No & 1D \\
		 Elastic-Net &   No & Yes & Yes & Yes & 2D \\
		 Fused-lasso &   Yes & Yes & Yes & Yes & 2D \\
		 OSCAR &   No & Yes & Yes & Yes & 2D \\
		 L1CP &   No & Yes & Yes & Yes & 2D \\
		 Clustered-Lasso &   No & Yes & Yes & Yes & 2D \\
		 DLSelect+Ridge &   No & Yes & Yes & Yes & 1D \\
		 \bottomrule
		\end{tabular}	
		\caption{Comparision Table}\label{table:block}
\end{table}
\end{small}


\section{Dual Lasso Selector}

In this section, we present the dual Lasso selector, a new variable selection method for sparse high-dim regression models with correlated variables. First, we study the theoretical properties of the Lasso and dual Lasso solutions. Then, we show that the magnitude of correlations between the predictors and the dual vector determines the set of active predictors. This is the basis for our correlated variable selection.

The dual problem of the Lasso problem (\ref{eq:lasso}) can be given as follows (we provide the detailed derivation of the Lasso's dual in the appendix A.1):

\begin{align}
	\sup_{\theta} \; &  \frac{1}{2} \|\textbf{Y} \|_2^2 - \| \theta - \textbf{Y} \|_2^2 \\
	\text{subject to } & | X_j^T \theta|\leq \lambda \text{ for } j = 1,...,p ,
\end{align}
where $\theta$ is the dual vector, as defined in equation (\ref{eq:lang}). For a fixed $\lambda \geq 0$, let $\hat{\beta}_{lasso}(\lambda)$ and $\hat{\theta}(\lambda)$ denote the optimal solutions of the Lasso and its dual problem respectively. Since it is implicit that the Lasso and its dual optimal depends on the $\lambda$, we drop the term $\lambda$ from the expression for notational simplicity. 
From KKT conditions (derivation of the KKT conditions is given in the appendix A.2) we get the following primal dual relationship:
\begin{align}
	\hat{\theta} = \textbf{Y} - \textbf{X} \hat{\beta}_{lasso}. \label{eq:primal_dual}
\end{align}

It is worth mentioning the basic properties of the Lasso and its dual, which has already been derived and studied by various authors (see \cite{tibshirani2011} and \cite{Jie} for more insights). 
\begin{enumerate}
\item \textbf{Uniqueness of the Lasso-fit}: There may not be a unique solution for the Lasso problem because for the criterion (\ref{eq:lasso_recall}) is not strictly convex in $\beta$. But the least square loss is strictly convex in $\textbf{X} \beta$, hence there is always a unique fitted value $\textbf{X} \hat{\beta}$. 

\item \textbf{Uniqueness of the dual vector}: The dual problem is strictly convex in $\theta$, therefore the dual optimal  $\hat{\theta}$ is unique. Another argument for the uniqueness of $\hat{\theta}$ is that it is a function of $\textbf{X} \hat{\beta}$ (\ref{eq:primal_dual}) which itself is unique.
 The fact that the DLS can achieve consistent variable selection for situations (with correlated active predictors) when the Lasso is unstable for estimation of the true active set is related to the uniqueness of the dual Lasso solution.

\item \textbf{Uniqueness of the Sub-gradient}: Sub-gradient of $\ell_1$ norm of any Lasso solution $\hat{\beta}$ is unique because it is a function of $\textbf{X} \hat{\beta}$ (see Appendix A.2).  More specifically, suppose $\hat{\beta}$ and $\tilde{\beta}$ are two lasso solutions for a fixed $\lambda$ value, then they must have the same signs $sign(\hat{\beta}) = sign(\tilde{\beta})$, it is not possible that $\hat{\beta}_j > 0 $ and $\hat{\beta}_j < 0 $ for some $j$.

\end{enumerate}
Let $\hat{S}_{lasso}$ denote the support set or active set of the Lasso estimator $\hat{\beta}$ which is given as 
 \begin{align}
 	\hat{S}_{lasso}(\lambda) = \{j \in \{ 1,...,p \} : (\hat{\beta}_{lasso})_{j} \neq 0\}
 \end{align}
Similarly, we define the active set of the dual Lasso vector that corresponds to the active constraints of the dual optimization problem. We note that constraints are said to be active at a feasible point if that point lies on a boundary formed by the constraint. 

 \begin{align}
 	\hat{S}_{dual}(\lambda) = \{j \in \{ 1,...,p \} : \; | X_j^T \theta| =  \lambda \}
 \end{align}

Now, we define the following lemmas that will be used later for our mathematical derivations.

\begin{lemma} \label{lemma:lasso_in_dual}
The active set selected by the Lasso $\hat{S}_{lasso}(\lambda)$ is always contained in the active set selected by the dual Lasso $\hat{S}_{dual}(\lambda)$, that is  
\[ \hat{S}_{lasso}(\lambda) \subseteq \hat{S}_{dual}(\lambda).
\] 
\end{lemma} 
\begin{proof}
The proof is rather easy. From KKT condition (dual feasibility condition, see Appendix A.2), we have 
\begin{align}
	 | X_j^T \theta| < \lambda \implies  \hat{\beta}_j = 0  \label{eq:inactive_feature}
\end{align}
 The proof lies in the ``implication" in the above equation (\ref{eq:inactive_feature})(but not in equivalence).
\end{proof}

It is known that Irrepresentable condition (assuming beta-min
conditions holds) is necessary and sufficient condition for the Lasso to select true model, see \cite{Zhao} (for completeness we have proved it in appendix A.4).

\begin{lemma} \label{lemma:lasso_dual}
Under assumption of the Irrepresentable Condition (IC) on the design matrix, the active set selected by the Lasso $\hat{S}_{lasso}(\lambda)$ is equal to the active set selected by the dual Lasso $\hat{S}_{dual}(\lambda)$, that is  
\[ \hat{S}_{lasso}(\lambda) = \hat{S}_{dual}(\lambda).
\] 
\end{lemma}
The proof is worked out in Appendix A.3.

The IC may fail to hold due to violation of any one of (or both) the following two conditions:
\begin{enumerate}
\item When $C_{11}$ is not invertible, that implies there is strong correlation among variables of the true active set.
\item The active predictors are correlated with the noise features (this situation is better explained in terms of irrepresentable condition).
\end{enumerate} 
When there is strong correlation among variables of the active set, then $C_{11}$ is not invertible and the IC does not hold, and the Lasso fails to do variable selection. But we argue that the dual Lasso can still perform variable selection consistently even when $C_{11}$ not invertible, when we impose some milder condition on the design matrix, we call it Pseudo Irrepresentable Condition (PIC). The Pseudo Irrepresentable Condition is defined as follows.

\begin{definition} [Pseudo Irrepresentable Condition(PIC)]
We partition the covariance matrix as in (\ref{eq:sigma_par}). Then the Pseudo Irrepresentable condition is said to be met for the set S with a constant $\eta >0 $, if the following holds:
\begin{align} \label{eq:PIC}
	|X^T_j G \;sign(\beta_1) | \leq 1 - \eta, \text{ for all } j \in S^c, 
\end{align}
where G is a generalized inverse of the form $\left[ \begin{array}{cc}
	C_{A}^{-1}  & 0 \\
	0 & 0 \end{array} \right] $, and  (\ref{eq:PIC}) holds for all $C_A \in C_R$, where $C_R$ is defined as $C_R := \{C_{rr}: rank(C_{rr}) = rank(C_{11})= r, C_{rr} \subset C_{11} \}$. 
\end{definition}

The following lemma gives the sufficient condition for the dual Lasso for support recovery. This lemma is similar in spirit of the Lemma 2 define in \cite{Omidiran}. Here, we do not assume that $\Sigma_{11}$ is invertible. 
\begin{lemma}[Primal-dual Condition for Variable Selection]
Suppose that we can find a primal-dual pair $(\hat{\beta}, \hat{\theta})$ that satisfy the KKT conditions 
\begin{align}
	\textbf{X}^T (\textbf{Y} - \textbf{X} \hat{\beta} )+ \lambda \hat{v} &= 0, \; \text{ where } \hat{v} = sign(\hat{\beta})\\
	\hat{\theta} &= \textbf{Y} - \textbf{X} \hat{\beta},
\end{align} 
and the signed support recovery conditions
\begin{align}
	\hat{v}_j &= sign(\beta_j) \text{ for all } j \in S, \\
	\hat{\beta}_j & = 0 \text{ for all } j \in S^c ,\\
	|\hat{v}_j| & < 1  \text{ for all } j \in S^c  \label{eq:dualSign}
\end{align}
Then $\hat{\theta}$ is the unique optimal solution to the dual Lasso and $\hat{S}_{dual}$ recovers the true active set. 
\end{lemma}
We have shown that the dual Lasso optimal  $\hat{\theta}$ is always unique, and it remains to show that the $\hat{S}_{dual}$ recovers the true active set. Under the assumption (\ref{eq:dualSign}), we can derive that $|X_j^T \hat{\theta}|  < \lambda $ for all $j \in S^c $. Therefore $\hat{S}_{dual} = S$.
 
\begin{theorem}
Under assumption of the PIC on the design matrix $\textbf{X}$, the active set selected by the dual Lasso $\hat{S}_{dual}$, is the same as the true active set $S$ with high probability. that is  
\[\hat{S}_{dual} = S.
\] 
\end{theorem}

When $C_{11}$ is invertible the PIC coincides with the IC, and under assumption of the IC we have already shown that $\hat{S}_{dual} = \hat{S}_{lasso}$. In Appendix A.4, we prove that the PIC is necessary and sufficient condition (beta-min condition is implicit) for the dual Lasso to consistently select the true active set. The PIC may hold even when $C_{11}$ is not invertible, which implies that the PIC is weaker than the IC. We illustrate it with the following examples:

Let $S = \{1,2,3,4 \}$ be the active set, the covariance matrix $C = \frac{\textbf{X}^T \textbf{X}}{n} $ and is given as
\begin{align*}
	C =  \left[ \begin{array}{ccccc}
	1  & 0  & 0 & 0 & \rho \\
   0   & 1  & 0 & 0 & \rho \\
   0   & 0  & 1 & 0 & \rho \\
   0   & 0  & 0 & 1	& \rho \\
 \rho & \rho & \rho& \rho & 1
	\end{array} \right],
\end{align*}
where the active variables are uncorrelated and the noise variable is equally correlated with all active covariates. 
First of all, it is easy to check that only for $ | \rho | \leq \frac{1}{2} $, $C$ is positive semi definite, and for $ | \rho | < \frac{1}{4}$, $C$ satisfies the IC.

Now, we augment this matrix with two additional columns, one copy of the first and second active variables, and we rearrange the columns such that we get the following covariance matrix, and we redefine the set of active variables as $S = \{1,2,3,4,5,6 \}$ and we assume that $ | \rho | < \frac{1}{4}$.

\begin{align*}
	C =  \left[ \begin{array}{ccccccccc}
	1 & 1  & 0 & 0 & 0 & 0 & \rho \\
	1 & 1  & 0 & 0 & 0 & 0 & \rho \\
   0 & 0   & 1 & 1  & 0 & 0 & \rho \\
   0 & 0   & 1 & 1  & 0 & 0 & \rho \\
   0 & 0 &  0   & 0  & 1 & 0 & \rho \\
   0 & 0 & 0   & 0  & 0 & 1	& \rho \\
 \rho & \rho & \rho & \rho & \rho& \rho & 1
	\end{array} \right].
\end{align*}
We partition  $\Sigma$ as (\ref{eq:sigma_par}), and it is clear that $C_{11}$ is not invertible and IC does not hold, hence the Lasso does not perform variable selection. The rank of the $C_{11}$ is 4. Let us consider any $(4 \times 4)$ sub matrix of $C_{11}$ such that its rank is four ($ S_1 \subset S, and rank(S_1)= 4, S_1 = \{ \{1,3,5,6 \}, \{1,4,5,6 \}, \{2,3,5,6 \}, \{2,4,5,6 \}$). Further, we partition $C_{11}$ as 
\begin{align} \label{eq:sigma11_par}
	C_{11} = \left[ \begin{array}{cc}
	C_{rr}  & C_{rr'}\\
	C_{r'r} & C_{r'r'}	
	\end{array} \right],
\end{align}
where $rank(C_{rr})$ has full column rank and $rank(C_{rr}) = rank(C_{11})$, let $C_R$ be set of are four such possible invertible sub matrices of $C_{11}$. Then considering the generalized inverses corresponding to them as
\begin{align} \label{eq:sigma11_inv}
	C_{11}^{+} = \left[ \begin{array}{cc}
	C_{A}^{-1}  & 0 \\
	0 & 0	
	\end{array} \right],
\end{align}
where $C_A \in C_R$ is invertible. With the above inverse $C_{11}^{+}$ the PIC holds for the design matrix $\textbf{X}$. It can be also viewed as the IC is satisfied for each reduced active set $S' \in S_1$ and the corresponding reduced design matrix $\textbf{X}_{S'}$, and hence the Lasso picks randomly one element from the set $S_1$ and sets the coefficient of the noise variable to zero (with high probability). Also, since PIC holds, the dual Lasso will select the true active set $S$ with high probability and will set zero for the coefficient of noise feature.

\subsection{Dual Lasso Selection and Ridge Estimation}
After proving that the joint consideration of the Lasso primal and its dual leads to correlated variable selection (under certain regularity condition), we now combine the dual Lasso selection with the Ridge estimation. Mainly, we consider the $\ell_2$ penalty (Ridge penalty) which is known to perform best in case of correlated variables, to estimate the coefficients of the predictors selected by the dual Lasso. We develop an algorithm called DLSelect+RR, which is a two stage procedure, the dual selection followed by the Ridge Regression.  

\begin{algorithm}[h!]
\SetAlgoLined
\textbf{Input:} dataset $(\textbf{Y},\textbf{X})$\\
\textbf{Output:} $\hat{S}$:= the set of selected variables\\
$\hat{\beta}$ := the estimated coefficient vector\\
\caption{DLSelect+RR Algorithm}\label{algo:DLSelect}
\textbf{Steps:}\\
  1. Perform Lasso on the data $(\textbf{Y}$, $\textbf{X})$.
  Denote the Lasso estimator as $\hat{\beta}_{lasso}$.\\
  2. Compute the dual optimal as 
  \[\hat{\theta} = \textbf{Y} - \textbf{X}\hat{\beta}_{lasso}. \]
  Denote the dual Lasso active set as $\hat{S}_{dual}$\\
  3. Compute the reduced design matrix as \[\textbf{X}_{red} = \{ {X}_j : j \in \hat{S}_{dual} \}. \]
  4. Perform Ridge regression based on the data  $(\textbf{Y}$, $\textbf{X}_{red})$ and obtain the ridge estimator $\beta_j$ for $j \in S_{dual} $. Set the remaining coefficients to zero.
		 \[ \hat{\beta}_j = 0 \text{ if } j \not\in S_{dual} \]
  \textbf{return} $(\hat{S}, \hat{\beta})$ 
\end{algorithm}

 If model selection works perfectly (under strong assumptions, i.e. IC), then the post-model selection estimators are the oracle estimators with well behaved properties (see \cite{Belloni}). In the following we argue that for the combination, dual selection followed by $\ell_2$ estimation, the prediction accuracy is at least as good as the Lasso.
 It has been already proven that the Lasso+OLS (\cite{Belloni}) estimator performs at least as good as Lasso in terms of the rate of convergence, and it has a smaller bias than the Lasso. Further Lasso+mLS (Lasso+ modified OLS) or Lasso+Ridge estimator have been also proven to be asymptotically unbiased under the Irrepresentable condition and other regularity conditions, see \cite{liu2013}. Under the Irrepresentable condition the Lasso solution is unique and the DLSelect+RR is the same as the Lasso+Ridge and the same argument holds for the DLSelect+RR. Also, In the following section we prove empirically that the prediction performance of the DLSelect+RR is at least as good as the Lasso.
 


\section{Numerical Studies}

In this section, we apply the DLSelect+Ridge for variable selection and estimation on simulations and real data and compare the results with that of the Lasso, Ridge and Elastic-net. We consider the True Positive Rate (power) and False Discovery Rate (FDR) as the measure of performances for variable selection, which are defined as follows.
\begin{equation}\label{eq:tpr}
\begin{aligned}  
	TPR &= \frac{ | \hat{S} \bigcap S  |} {|S|}\\
	FDR & = \frac{ | \hat{S} \bigcap S^c  |} {|\hat{S}|}
\end{aligned}
\end{equation}
For prediction performance we consider the Mean Squared Prediction Error, which is defined as 
\begin{align}
	 MSE &= \frac{1}{n}\|\textbf{Y} - \hat{\textbf{Y}}\|^2_2,
\end{align}
where $\hat{\textbf{Y}}$ is the predicted response vector or an estimate $\textbf{X} \hat{\beta}$ based on an estimator $\hat{\beta}$.

Since our aim is to avoid false negatives, we do not report false positives, and ridge does not perform variable selection therefore TPR is not reported for the Ridge. The Ridge is considered as a competitor because its prediction performance is better than the Lasso for correlated designs.

\subsection{Simulation Examples} 
We consider five different simulation settings, 
where simulate data from the linear model as in (\ref{eq:lr}) with fixed design matrix $\textbf{X}$, and $\sigma = 1$.  We generate the design matrix $\textbf{X}$ once from a multivariate normal distribution $N_p(0, \Sigma)$ with different structures for $\Sigma$, and keep it fixed for all replications.

For each simulation example, $100$ data sets were generated, where each dataset consists of  a training set used to fit the model, an independent validation set used for tuning the regularization parameter and an independent test set used for evaluation of the performance. We denote by $ \#/\#/\#$, the number of observation in  training, validation and test set respectively. 
For most of the simulation examples we fix the size of the active set to $s=20$ and the true coefficient vector as 
\begin{align} \label{eq:true_beta}
	\beta = \{ \underbrace{1,...,1}_{20},  \underbrace{0,...,0}_{480} \}.
\end{align} 

We generate $100$ data sets with sample sizes $n/n/1000$ with $n = 100, 200, 400 , 600$. For each simulation example and each method the MSE and TPR are computed over $100$ data sets.  A suitable grid of values for the tuning parameters is considered, and all reported results are based on the median of $100$ simulation runs.

\subsubsection{Block Diagonal Model}
Here we generate the fixed design matrix $\textbf{X} \sim N_p(0, \Sigma_1)$ with $p=500$, where $\Sigma_1$ is a block diagonal matrix. The matrix $\Sigma_1$ consists of $50$ independent blocks $B$ of size $10 \times 10$, defined as
\[
	B_{j,k} =  \left\lbrace \begin{array}{cc}
	1, & j=k\\
	.9, & otherwise	
	\end{array} \right. \]
 
This simulation example is considered to show that when the Lasso (due to collinearity) and Ridge (due to noise) do not perform well, the Enet and DLSelect+Lasso perform quite well. 
From the table (\ref{table:BD}) it is easy to figure out that the Ridge performs poorly in terms of prediction performance for all simulation setting and the Lasso is not stable for variable selection. The Enet consistently selects the true active set, and DLSelect+Ridge completes with Enet in all settings.
\begin{table}[h!] \centering
		\caption{Performance measures for block diagonal case}\label{table:BD}
		\begin{tabular}{ p{10mm}p{30mm}p{25mm}p{25mm}}
		\toprule 
		 n & Method& MSE(SE) & TPR \\
		\toprule 
		100 & Lasso & 22.37(1.31) &  0.45 \\	
		& Ridge &  565.58(3.31) & NA \\	
		& Enet& 22.17(1.2) & 1  \\
		& DLSelect+Ridge & 18.92(1.2) & 0.6 \\
		
		200 & Lasso & 11.52(0.67) &   0.6 \\	
		& Ridge &  466.35(2.26) & NA \\	
		& Enet& 11.37(0.63) & 1  \\
		& DLSelect+Ridge & 8.45(0.55) & 1 \\
		
		400 & Lasso & 6.88(0.31)& 0.55 \\	
		& Ridge &  417.59(2.07) & NA \\	
		& Enet& 6.85(0.32)& 1  \\
		& DLSelect+Ridge & 5.42(0.37) & 1 \\
		
		600 & Lasso & 5.54(0.28) & 0.65 \\	
		& Ridge &  5.87(0.31) & NA \\	
		& Enet& 5.34(0.25)& 1  \\
		& DLSelect+Ridge & 3.53(0.22) & 1 \\
		\bottomrule
		\end{tabular}	
\end{table}

\subsubsection{Single Block Model with Noise Features}
Here we generate the fixed design matrix $\textbf{X} \sim N_p(0, \Sigma_2)$ with $p=500$, where $\Sigma_2$ is almost an identity matrix except for the first $20 \times 20$ is a single highly correlated block. The matrix $\Sigma_2$ is defined as
\[
	\Sigma_{j,k} =  \left\lbrace \begin{array}{cc}
	1, & j=k\\
	.9, & j \neq k \textbf{ and } i,j \leq 20 \\	
	0, & otherwise
	\end{array} \right. \]
In this setting, the first twenty variables are active predictors and they are highly correlated, and the remaining $480$ are independent noise variables. We generate $100$ data sets with sample sizes $n/n/1000$, where $n = 100, 200, 400 , 600$. The simulation results are reported in Table (\ref{table:SBN}).
   
\begin{table}[h!] \centering
		\caption{Performance measures for single block with noise }\label{table:SBN}
		\begin{tabular}{ p{10mm}p{30mm}p{25mm}p{25mm}}
		\toprule 
		 n & Method& MSE(SE) & TPR \\
		\toprule 
		100 & Lasso &  101.37(2.26) &  0.4 \\	
		& Ridge &  922.41(4.36) & NA \\	
		& Enet& 102.91(2.37) & 1  \\
		& DLSelect+Ridge & 88.00(3.2) & 1 \\
		
		200 & Lasso & 15.55(0.59) &   0.25 \\	
		& Ridge &   627.66(2.67) & NA \\	
		& Enet& 15.92(0.57) & 1  \\
		& DLSelect+Ridge & 8.26(0.56) & 1 \\
			
		400 & Lasso & 2.67(0.17) & 0.2 \\	
		& Ridge &  456.17(1.95) & NA \\	
		& Enet& 2.50(0.15) & 1  \\
		& DLSelect+Ridge & 1.16(0.081) & 1 \\
		
		600 & Lasso & 1.46(0.06) & 0.15 \\	
		& Ridge &  5.87(0.26) & NA \\	
		& Enet& 1.12(0.06) & 1  \\
		& DLSelect+Ridge &  2.29(0.13) & 1 \\
		\bottomrule
		\end{tabular}	
\end{table}
From Table (\ref{table:SBN}), it is clear that the Lasso and Ridge performs poorly (one can give similar argument as Block diagonal model). The Enet and DLSelect+Ridge consistently selects true active set with reduced prediction error.
\subsubsection{Single Block Model without Noise Features}
Here we generate the fixed design matrix $\textbf{X} \sim N_p(0, \Sigma_3)$ with $p=20$, where $\Sigma_3$ is a single block of highly correlated variables. The matrix $\Sigma_3$ is defined as
\[
	\Sigma_{j,k} =  \left\lbrace \begin{array}{cc}
	1, & j=k\\	
	.99, & otherwise
	\end{array} \right. \]
The true coefficient vector is \[\beta = \{ \underbrace{1,...,1}_{20} \}. \]
We generate $100$ data sets with sample sizes $n/n/200$ with $n = 20, 200 $. The simulation results are reported in Table (\ref{table:SBWN}).
	
	\begin{table}[h!] \centering
		\caption{Performance measures for single block with noise }\label{table:SBWN}
		\begin{tabular}{ p{10mm}p{30mm}p{25mm}p{25mm}}
		\toprule 
		 n & Method& MSE(SE) & TPR \\
		\toprule 
				 
		20 & Lasso & 245.95(8.91) &   0.25 \\	
		& Ridge &   246.23(8.97) & NA \\	
		& Enet& 244.95(8.85) & 1  \\
		& DLSelect+Ridge & 249.99(7.15) & 1 \\
		
		200 & Lasso &  4.75(0.42) &  0.1 \\	
		& Ridge &  4.64(0.43) & NA \\	
		& Enet& 4.75(0.40) & 1  \\
		& DLSelect+Ridge & 1.00(0.10) & 1 \\
		
		\bottomrule
		\end{tabular}	
\end{table}
From Table (\ref{table:SBWN}), it is apparent that the Lasso performs poorly in terms of variable selection as well as prediction accuracy. The Ridge gives the best predictive performs for the number of sample size increases.
The Enet and DLSelect+Ridge consistently selects true active set with, however the DLSelect+Ridge has better prediction accuracy for moderate sample size.

\subsubsection{Toeplitz Model}
Here we consider special case of a Toeplitz matrix $\Sigma_4$ to generate the fixed design matrix $\textbf{X} \sim N_p(0, \Sigma_4)$ with $p=500$. The matrix $\Sigma_4$ is defined as
\[
	\Sigma_{j,k} =  \left\lbrace \begin{array}{cc}
	1, & j=k\\
	\rho^{|i-j |} & otherwise
	\end{array} \right. \], where $\rho = 0.9$.
The true coefficient vector is as defined in (\ref{eq:true_beta}), and we generate $100$ data sets with sample sizes $n/n/1000$ where $n = 100, 200, 400 , 600$. The Table \ref{table:TP} shows the simulation results.

\begin{table}[h!] \centering
		\caption{Performance measures for Toeplitz settings}\label{table:TP}
		\begin{tabular}{ p{10mm}p{30mm}p{25mm}p{25mm}}
		\toprule 
		 n & Method& MSE(SE) & TPR \\
		\toprule 
		100 & Lasso & 10440.42(53.28) &  0.36 \\	
		& Ridge &  12478.77(23.98) & NA \\	
		& Enet& 10352.06(23.21) & 0.89  \\
		& DLSelect+Ridge &  7789.103(22.61) & 1\\
		
		200 & Lasso & 713.34(4.13) &   0.51 \\	
		& Ridge & 654.85(3.8) & NA \\	
		& Enet&  651.08(3.86) & 0.99  \\
		& DLSelect+Ridge & 97.17(1.70) & 0.77 \\
	
		400 & Lasso & 145.15(2.12)& 0.54 \\	
		& Ridge &  98.63(1.13) & NA \\	
		& Enet& 103.57(1.19) & 1  \\
		& DLSelect+Ridge & 52.19(0.85) & 1 \\

		600 & Lasso & 200.15(1.68) & 0.65 \\	
		& Ridge &  169.01(1.27) & NA \\	
		& Enet& 169.66(1.31)& 0.99  \\
		& DLSelect+Ridge & 22.35(0.47)) & 1 \\
		\bottomrule
		\end{tabular}	
\end{table}

The Table (\ref{table:SBWN}) shows that the Lasso and the Ridge performs poorly for all settings. The DLSelect+Ridge consistently selects the true active set, however the DLSelect+Ridge has better prediction accuracy for moderate sample size.

\subsubsection{Independent Predictor Model}
Finally we consider an identity matrix to generate  the fixed design matrix $\textbf{X} \sim N_p(0, I)$ with $p=500$. In this setting all predictors and uncorrelated. The true coefficient vector is as defined in (\ref{eq:true_beta}), and we generate $100$ data sets with sample sizes $n/n/1000$, where $n = 100, 200, 400 , 600$. 

The Table \ref{table:Indep} shows the simulation results.

\begin{table}[h!] \centering
		\caption{Performance measures for independent settings}\label{table:Indep}
		\begin{tabular}{ p{10mm}p{30mm}p{25mm}p{25mm}}
		\toprule 
		 n & Method& MSE(SE) & TPR \\
		\toprule 
		100 & Lasso & 14.50(3.74) &  1 \\	
		& Ridge &  153.47(0.91) & NA \\	
		& Enet&  28.85(4.59) & 1  \\
		& DLSelect+Ridge & 39.16(29.73) & 1 \\
		
		200 & Lasso & 2.32(0.23) &   1 \\	
		& Ridge & 139.34(0.76) & NA \\	
		& Enet&  2.45(0.24) & 1  \\
		& DLSelect+Ridge & 3.48(0.51) & 1 \\
		
		400 & Lasso & 1.56(0.10) & 1 \\	
		& Ridge &  118.21(0.73)  & NA \\	
		& Enet& 1.59(0.10) & 1  \\
		& DLSelect+Ridge & 3.70(0.46) & 1 \\
		
		600 & Lasso & 1.35(0.06) & 1 \\	
		& Ridge &  8.60(0.53) & NA \\	
		& Enet& 1.37(0.06)& 1  \\
		& DLSelect+Ridge & 3.37(0.35) & 1 \\
		\bottomrule
		\end{tabular}	
\end{table}

The Table (\ref{table:SBWN}) shows that the Lasso gives the best prediction accuracy and the Ridge performs poorly for all the settings. The Enet and DLSelect+Ridge competes each other.

\subsection{Real Data Example}
In this section, we consider five real world data to evaluate the prediction and variable selection performance of the proposed method $DLSelect+Ridge$. We randomly split the data sets into two halves for $100$ times, we use first half for training (using cross validation) and second half is used as a test set. For testing variable selection, For first two datasets (UScrime and Prostate) we consider all the variables as relevant variables and for the remaining datasets we select ten most variable which are highly correlated with the response and another ten variables which are correlated with the selected variables. Median MSE, standard error and median TPR are reported over $100$ splits for each example.

\subsubsection{USCrime Data}
This is a classical  dataset collected in 1960 where criminologists are mainly interested in the effect of punishment on crime rates. 
 There are Independent $15$ independent variables and the response is rate of crimes in a particular category per head of population. For more details on this dataset we refer to \cite{UScrime}. The performance measures are reported in Table \ref{table:USCrime}.
\begin{table}[h!] \centering
		\caption{Performance measures for UScrime data}\label{table:USCrime}
		\begin{tabular}{ p{10mm}p{30mm}p{25mm}p{25mm}}
		\toprule 
		  & Method& MSE(SE) & TPR \\
		\toprule 
		& Lasso & 87725(36371) &  0.54 \\	
		& Ridge &  77153(26118) & NA \\	
		& Enet&  83403(34342) & 0.45  \\
		& DLSelect+Ridge & 78275(24625) & 0.54 \\
		\bottomrule
		\end{tabular}	
\end{table}

Here, we have considered all covariates as important variables. The Ridge regression outperforms the other methods, and DLSelect+Ridge performs better than Lasso and the Enet in terms of prediction perform as well as variable selection.
\subsubsection{Prostate Data}
The Prostate dataset has $97$ observations and $9$ covariates.
This dataset is an outcome of a study that examined the correlation between the level of prostate specific antigen and a number of clinical measures in men who were about to receive a radical prostatectomy. For further details on the dataset we refer to \cite{Prostate}.
\begin{table}[h!] \centering
		\caption{Performance measures for Prostate data}\label{table:Indep}
		\begin{tabular}{ p{10mm}p{30mm}p{25mm}p{25mm}}
		\toprule 
		  & Method& MSE(SE) & TPR \\
		\toprule 
		& Lasso &  0.56(0.09) &  0.63 \\	
		& Ridge &  0.56(0,08) & NA \\	
		& Enet&  0.55(0.09) & 0.63  \\
		& DLSelect+Ridge & 0.56(0.07) & 1\\
		\bottomrule
		\end{tabular}	
\end{table}
The performance measures are reported in Table \ref{table:USCrime}. Here, we have considered all covariates as important variables, From the table, it is clear that all method seems to report almost the same prediction error, and DLSelect+Ridge performs better than Lasso and the Enet in terms of variable selection.

\subsubsection{Riboflavin Data}
The dataset of riboflavin consists of, $n=71$ observations of $p=4088$ predictors (gene expressions) and univariate response, riboflavin production rate(log-transformed), see \cite{HDview} for details on riboflavin dataset. Since the ground truth is not available, we consider Riboflavin data for the design matrix $\textbf{X}$ with synthetic parameters $\beta$ and simulated Gaussian errors $\epsilon \sim \mathbb{N}_n(0, \sigma^2 I)$. We fix the size of the active set to $s = 20$ and $\sigma = 1$ and for the true active set, select ten predictors which are highly correlated with the response and another ten variables which are most correlated with those selected variables. The true coefficient vector is 
\begin{align*}
 \beta_j = \left\lbrace \begin{array}{ll}
	1 & \text{ if } j \in S  \\
	0 & \text{ if } j \not\in S\\
\end{array}	 \right. .
\end{align*}
 Then we compute the response using the Equation (\ref{eq:lr}).
The performance measures are reported in Table  \ref{table:Ribo}.
\begin{table}[h!] \centering
		\caption{Performance measures for Leukaemia data}\label{table:Ribo}
		\begin{tabular}{ p{10mm}p{30mm}p{25mm}p{25mm}}
		\toprule 
		  & Method& MSE(SE) & TPR \\
		\toprule 
		& Lasso &   96.69(63) &  0.27 \\	
		& Ridge &  290.98(138) & NA \\	
		& Enet&   92.44(65) & 0.44  \\
		& DLSelect+Ridge & 88.31(54) & 0.38\\
		\bottomrule
		\end{tabular}	
\end{table}
From the table (\ref{table:Ribo}), we conclude that Enet outperforms in terms of variable selection, whereas, DLSelect+Ridge performs better than others in terms of prediction performance.
\subsubsection{Myeloma Data}
We consider another real dataset, Myeloma $(n = 173, p = 12625)$ data for the design matrix $\textbf{X}$ with synthetic parameters $\beta$ and simulated Gaussian errors
We refer to \cite{myeloma} for details on Myeloma dataset. In this example also, we set active set and generated response same as previous example (Riboflavin). The performance measures are reported in Table \ref{table:Myeloma}.

\begin{table}[h!] \centering
		\caption{Performance measures for Leukaemia data}\label{table:Indep}
		\begin{tabular}{ p{10mm}p{30mm}p{25mm}p{25mm}}
		\toprule 
		  & Method& MSE(SE) & TPR \\
		\toprule 
		& Lasso &  68.29(27.73) &  0.35 \\	
		& Ridge & 239.58(58.25) & NA \\	
		& Enet&   70.37(29.16) & 0.58  \\
		& DLSelect+Ridge & 75.25(28.70) & 0.52\\
		\bottomrule
		\end{tabular}	
\end{table}

From the table (\ref{table:Ribo}), the Enet outperforms in terms of variable selection as well as prediction performance.

\subsubsection{Leukaemia Data}
We consider the famous dataset of Leukaemia Data \cite{leuk}. In this example also, we set active set and generated response same as previous examples. The performance measures are reported in Table \ref{table:leuk}.
\begin{table}[h!] \centering
		\caption{Performance measures for Leukaemia data}\label{table:leuk}
		\begin{tabular}{ p{10mm}p{30mm}p{25mm}p{25mm}}
		\toprule 
		  & Method& MSE(SE) & TPR \\
		\toprule 
		& Lasso &   111.53(97.3) &  0.46 \\	
		& Ridge &  182.11(97.2) & NA \\	
		& Enet&  90.47(82.64) & 0.6  \\
		& DLSelect+Ridge & 72.43(67.3) & 0.5\\
		\bottomrule
		\end{tabular}	
\end{table}
From the table (\ref{table:leuk}), it is clear that the Enet gives the better prediction performance, and DLSelect+Ridge performs better than Lasso and the Enet in terms of variable selection.


\section*{Appendix A}
\subsection*{A.1 Derivation of the Dual Form of the Lasso}
In this section, we derive the Lagrange dual of the Lasso problem (\ref{eq:lasso}), which serves as the selection operator for our approach. That is, by considering the lasso and its dual simultaneously it is possible to identify the non-zero entries in the estimator. For more details on dual derivation and projection on polytope formed by the dual constraints, we refer to \cite{Jie}.

We recall that the Lasso problem is defined as the following convex optimization problem.
\begin{align}
	 \min_{\beta \in \mathbb{R}^p} \left\lbrace \frac{1}{2}\|{ \textbf{Y} -\textbf{X} \beta}\|_2^2 + \lambda \|\beta\|_1 \right\rbrace \label{eq:lasso_recall}
\end{align}

Since the above problem has no constraints, its dual problem is trivial. So we introduce a new vector $\textbf{r} = \textbf{Y} -\textbf{X} \beta$, then the Lasso problem can be written as:

\begin{align}
	  \min_{\beta \in \mathbb{R}^p} & \left\lbrace \frac{1}{2}\| \textbf{r} \|_2^2 + \lambda \|\beta\|_1 \right\rbrace  \\
	\text{subject to } & \textbf{r} = \textbf{Y} -\textbf{X} \beta \nonumber
\end{align}

Now, to account for the constraints we introduce the dual vector $\theta in \mathbb{R}^n$, then we get the following Lagrangian equation with $\beta$ and $r$ as primal variables.

\begin{align}
	  L(\beta, \textbf{r}, \theta) =  \frac{1}{2}\| \textbf{r} \|_2^2 + \lambda \|\beta\|_1  + \theta^T(\textbf{Y} - \textbf{X}\beta - \textbf{r}) \label{eq:lang}
\end{align}

Then the dual function can be written as:
\begin{align*}
	 g(\theta) & = \inf_{\beta, \textbf{r}} L(\beta, \textbf{r}, \theta) \\
	  &= \frac{1}{2}\|\textbf{r}\|_2^2 + \lambda \|\beta\|_1  + \theta^T(\textbf{Y} - \textbf{X}\beta - z)\\
	  & = \theta^T y + \inf_{r} \left\lbrace  \frac{1}{2}\| \textbf{r} \|_2^2  - \theta^T \textbf{r} \right\rbrace  +\inf_{\beta} \left\lbrace   \lambda \|\beta\|_1  - \theta^T  \textbf{X}\beta \right\rbrace \\
	  & =  \theta^T y + \inf_{r} L_1(r)+\inf_{\beta} L_2(\beta)
\end{align*}
After solving the first optimization problem, we get
\begin{align}
	\inf_{r} L_1(r) = - \frac{1}{2} \| \textbf{r} \|_2^2 \label{eq:residual}
\end{align}

Since  $L_1(r)$ is non-differentiable, we consider its subgradient
\begin{align*}
	\partial L_1(\beta) = \lambda v -\textbf{X}^T \theta ,
\end{align*}
where v is the subgradient of $\|\beta\|_1$, and it satisfies $\|v \|_{\infty} \leq 1$ and $v^{T}\beta = \|\beta\|_1$. For $L_1$ to attain an optimum, the following must hold.

\begin{align*}
	\lambda v -\textbf{X}^T \theta & = 0\\
	\implies \textbf{X}^T \theta & = \lambda v 
\end{align*}
\begin{align}
	\therefore |X_j^T \theta| \leq \lambda \; \text{ for all } j \in \{ 1,...,p \} \label{eq:dual_con}
\end{align}
From (\ref{eq:residual}) and (\ref{eq:dual_con}), we get the dual objective function as:
\begin{align*}
	 g(\theta) &= \theta^T \textbf{Y} - \frac{1}{2} \theta^T\theta \\
  g(\theta) & = \frac{1}{2} \|\textbf{Y} \|_2^2 - \| \theta - \textbf{Y} \|_2^2
\end{align*}
Then the dual problem is given as:
\begin{align} \label{eq:dual_lasso}
	\sup_{\theta} \; & g(\theta) =  \frac{1}{2} \|\textbf{Y} \|_2^2 - \| \theta - \textbf{Y} \|_2^2 \nonumber \\
	\text{subject to } & \therefore |X_j^T \theta| \leq \lambda \; \text{ for all } j \in \{ 1,...,p \} 
\end{align}

\subsection*{A.2 Relationship Between the Lasso and its Dual Optimal}
In this section, we derive the relationship between the Lasso optimal and its dual optimal.

For a fixed $\lambda$, the Lasso problem (\ref{eq:lasso}) is convex in $\beta$ and it is strictly feasible since it has no constraints, therefore by Slater’s condition, strong duality holds. Let us suppose that $\hat{\beta}, \hat{r}$ and $\hat{\theta} $ are optimal primal and dual variables, then by the KKT conditions the following must hold.

\begin{align}
 0 & \in \partial_\beta L(\hat{\beta}, \hat{r}, \hat{\theta}) \label{eq:par_beta} \\
 \Delta_z  L(\hat{\beta}, \hat{r}, \hat{\theta}) & = \hat{r} - \hat{\theta }= 0 \label{eq:par_z}\\
  \Delta_\theta  L(\hat{\beta}, \hat{r}, \hat{\theta}) & = \textbf{Y} - \textbf{X}\hat{\beta }- \hat{r} = 0  \label{eq:par_theta}
\end{align}
From (\ref{eq:par_beta}) we get
\begin{align*}
 x^T \hat{\theta} & = \lambda \hat{v} \\
|X_j^T \theta| & \leq \lambda \; \text{ for all } j \in \{ 1,...,p \}
 \end{align*}
Or equivalently for all $j \in \{ 1,...,p \}$ the following must hold.
 \begin{align}
 X_j^T \hat{\theta}  =  \left\lbrace \begin{array}{ll}
	\lambda & \text{ if } \hat{\beta}  > 0 \\
	\in \left[ -\lambda, \lambda \right]    &  \text{ if } \hat{\beta}=0 \\
	-\lambda & \text{ if } \hat{\beta}  < 0   \\
\end{array}	 \right. \label{eq:dual_feasibility}
\end{align}
From the above equation (\ref{eq:dual_feasibility}), we get the following important result.
\begin{align}
	| X_j^T \hat{\theta} | < \lambda \implies  \hat{\beta}=0 \label{eq:lasso_in_dual}
\end{align}
Finally, from (\ref{eq:par_z}) and (\ref{eq:par_theta}) we get the following equality.
\begin{align} \label{dual_lasso_fit}
	\hat{\theta} = \textbf{Y} - \textbf{X} \hat{\beta}
\end{align}
and substituting value of $\hat{\theta}$ in (\ref{eq:dual_feasibility}) we get the following expression.
\begin{align} \label{eq:subgrad_lasso_fit}
	\textbf{X}^T(\textbf{Y} - \textbf{X} \hat{\beta}) = \lambda v.
\end{align}

\section*{A.3 Proof of Lemma \ref{lemma:lasso_dual}}
\begin{proof}
Without loss of generality we can assume that the first $s = |S|$ variables are the active variables, and we partition the empirical covariance matrix as in Equation (\ref{eq:sigma_par}), $\hat{\beta }= (\beta_1 \; \beta_2)^T $ and $\hat{ v }= (v_1 \; v_2)^T $ accordingly. Let us recall the IC (for the noiseless case for simplicity), it is defined as follows.
\begin{definition} [Irrepresentable Condition(IC)]
The irrepresentable condition is said to be met for the set S with a constant $\eta >0 $, if the following holds:
\begin{align} \label{eq:IC}
	\|C_{12} C_{11}^{-1}sign(\beta_1) \|_{\infty} \leq 1 - \eta.
\end{align}
\end{definition}
Under IC, the lasso solution is unique. If we further assume the beta-min condition then the following holds, see (\cite{Buhlmann1}) for the detailed proof. 
\[ S = \hat{S}_{lasso}.
\] 
The proof of the proposition (\ref{lemma:lasso_dual}) is fairly simple, we prove it by contradiction. Let us assume that $\hat{S}_{lasso} ! = \hat{S}_{dual}$, then from Proposition (\ref{pro:lasso_in_dual}) the Lasso active set  $\hat{S}_{lasso}$ is a proper subset of the dual active set $\hat{S}_{dual}$ , and it follows that there exists some $j \in \hat{S}^c$ for which the following condition is satisfied.
\begin{align*}
\beta_j = 0  \text{ and } |X_j^{T} \hat{\theta}| = \lambda
\end{align*}
Substituting value of $\hat{\theta} = \textbf{Y} - \textbf{X} \hat{\beta}$ (see Appendix A.2) and $\textbf{Y} =  \textbf{X} \beta$, we get the following.
\begin{align*}
	|X_j^{T} (\textbf{Y} - \textbf{X}\hat{\beta})| &= \lambda \\
	\implies |X_j^{T} \textbf{X} (\beta - \hat{\beta})| &= \lambda
\end{align*}
Under IC the Lasso selects the active sets,  so we have $\beta_2 = \hat{\beta}_2 = 0$,  some algebraic simplification gives the following equality.
\begin{align}
	\implies |(C_{21})_j (\beta_1 - \hat{\beta}_1)| = \lambda \label{eq:c_21}
\end{align}

From the KKT condition (see AppendixA.2) we have:
\begin{align*}
	X^T (\textbf{Y} - \textbf{X} \hat{\beta} )+ \lambda \hat{v} = 0 
\end{align*} 
where $\| v \|_{\infty} \leq 1$ and $\beta v = \| \beta \|_1 $. 
, by substituting $\textbf{Y} = \textbf{X}\beta$, we get
\begin{align*}
	\textbf{X}^T \textbf{X} (\hat{\beta }- \beta) &= - \lambda \hat{v }
\end{align*}
We can write the above equation in terms of partitions of $C = \hat{\Sigma}$ as follows.
\begin{align*}
\left[ \begin{array}{cc}
	C_{11} & C_{12}\\
	C_{21} & C_{22})	
	\end{array} \right] 	 ( \begin{array}{c} \beta_1- \hat{\beta_{1}} \\ \beta_2- \hat{\beta_{2}} \end{array}) & = \lambda ( \begin{array}{c} v_1 \\ v_2 \end{array})
\end{align*}
Since $\beta_2 = \hat{\beta_{2}}=0$, therefore, we get the following equality. 
\begin{align*}
\beta_1- \hat{\beta_{1}} = \lambda C_{11}^{-1} sign(\beta_1)
\end{align*}
Substituting value of $\beta_1- \hat{\beta_{1}}$ into the equation
(\ref{eq:c_21}) we get
\begin{align}
	 |(C_{21})_j \lambda C_{11}^{-1} sign(\beta_1)| = \lambda \\
	 |(C_{21})_j  C_{11}^{-1} sign(\beta_1)| = 1
\end{align}
It violet the IC, hence under assumption of IC,
\[ \beta_j = 0  \implies |X_j^{T} \hat{\theta}| < \lambda .\] Therefore  the following equality must hold, that completes the proof.
\[ \hat{S}_{lasso}(\lambda) = \hat{S}_{dual}(\lambda).
\] 
\end{proof}

\section*{Appendix A.4 IC implies Lasso Variable Selection}
\begin{proof}
This result and proof are from \cite{Buhlmann1}. The IC depends on the covariance of the predictors $C = \hat{\Sigma}$ and the signs of the unknown true parameter $\beta$ (beta-min condition is implicit). For simplicity, we prove it for the noiseless case, where $\textbf{Y} = \textbf{X}\beta$. We first assume that the IC holds and we will show that Lasso correctly identifies the active set $S$. From KKT condition as in (\ref{eq:subgrad_lasso_fit}), and substituting $\textbf{Y} = \textbf{X}\beta$, we get
\begin{align*}
	\textbf{X}^T \textbf{X} (\hat{\beta }- \beta) &= - \lambda v \\
	 \left[ \begin{array}{cc}
	C_{11} & C_{12}\\
	C_{21} & C_{22})	
	\end{array} \right] 	 ( \begin{array}{c} \beta_1- \hat{\beta_{1}} \\ \beta_2- \hat{\beta_{2}} \end{array}) & = \lambda ( \begin{array}{c} v_1 \\ v_2 \end{array})
\end{align*}
We note that, for the true parameter vector, $\beta_{2} $ is a null vector by definition. We get the following two equations after some simplification:
\begin{align} \label{eq:kkt_1}
	C_{11}(\beta_{1} - \hat{\beta_1}) - C_{12} \hat{\beta_2} &= \lambda v_1 \\
	C_{21}(\beta_{1} - \hat{\beta_1}) - C_{22} \hat{\beta_2} &= \lambda v_2
\end{align}
After some algebraic simplification of the first equation we get
\begin{align*}
	\hat{\beta_1}-\beta_{1} = C^{-1}_{11} (  C_{12} \hat{\beta_2} +\lambda v_1)\\
	\end{align*}
	Substituting value of $\hat{\beta_1}-\beta_{1}$, in the second equation 
	\begin{align*}
	C_{21} C^{-1}_{11} (  C_{12} \hat{ \beta_2 } +\lambda v_1) - \Sigma_{22} \beta_2 &= \lambda v_2 
	\end{align*}
	by multiplying both the sides with $\hat{\beta}_2^T$
	\begin{align*}
	\hat{\beta}_2^T (C_{22} - C_{21} C^{-1}_{11} C_{12} ) \hat{ \beta_2 } &= -\lambda \| \beta_2 \|_1 - \hat{ \beta_2 }^T C_{21} C^{-1}_{11} \lambda v_1
	\end{align*}
Applying holders inequality for the term $\hat{ \beta_2 }^T C_{21} C^{-1}_{11}\lambda v_1$ on RHS we get
\begin{align*}
	\hat{ \beta_2 }^T C_{21} C^{-1}_{11} \lambda v_1 & \leq \lambda \| \hat{ \beta_2 } \|_1 C_{21} C^{-1}_{11} v_1 \|_{\infty} \\
	\implies & \leq  \lambda \| \hat{ \beta_2 } \|_1.
\end{align*}
We get the following expression after substitution,
\begin{align*}
	\hat{\beta}_2^T (C_{22} - C_{21} C^{-1}_{11} C_{12} ) \hat{ \beta_2 } \leq - \lambda \| \hat{ \beta_2 } \|_1.
\end{align*}
	
Since $\lambda \hat{ \beta_2 } \|_1 > 0 $ we get the following inequality, 
\begin{align*}
	\hat{\beta}_2^T (C_{22} - C_{21} C^{-1}_{11} C_{12} ) \hat{ \beta_2 } \leq 0
\end{align*}
The matrix $(C_{22} - C_{21} C^{-1}_{11} C_{12} )$ is a positive semi-definite, we have arrived at a contradiction. Therefore $\hat{\beta}_{S^c} = 0$, for any Lasso solution it is true. Hence the Lasso correctly identifies all the zero components, and $\hat{S}_{lasso} \subset S$. 

Now, we assume that lasso selects the true active set, and we will show that the IC holds. Basically, It is given that $\hat{\beta}_2 = \hat{\beta}_{S^c} = 0$. Using the KKT condition again, and substituting $\hat{\beta}_2 = 0$ in (\ref{eq:kkt_1}), we get the following expression.
\begin{align*} 
	C_{11}(\beta_{1} - \hat{\beta_1}) &= \lambda v_1 \\
	C_{21}(\beta_{1} - \hat{\beta_1}) &= \lambda v_2
\end{align*}
After solving the above we have
\begin{align*} 
	C_{21} C_{11}^{-1} \lambda v_1 &= \lambda v_2
\end{align*}
Since  $\| v2 \|_{\infty} < 1$ and we have the following inequality.
\begin{align*} 
	\| C_{21} C_{11}^{-1} \lambda sign(\beta_1) \|_{\infty} & <  \lambda \\
	\implies \| C_{21} C_{11}^{-1} sign(\beta_1) \|_{\infty} <  1
\end{align*}
\end{proof}

\section*{Appendix A.5 PIC implies dual Lasso Variable selection}
\begin{proof}
The proof is similar to the proof given for IC (see Appendix A.4) except we replace $G_{11}^{-1}$ with the one of the generalized inverse . The PIC, like IC depends on the covariance of the predictors $C = \hat{\Sigma}$ and the signs of the unknown true parameter $\beta$. For simplicity, we prove it for the noiseless case, where $\textbf{Y} = \textbf{X}\beta$. We first assume that the PIC holds and we will show that dual Lasso correctly identifies the active set $S$. From KKT condition as in (\ref{eq:subgrad_lasso_fit}), and substituting $\textbf{Y} = \textbf{X}\beta$, we get
\begin{align*}
	\textbf{X}^T \textbf{X} (\hat{\beta }- \beta) &= - \lambda v \\
	 \left[ \begin{array}{cc}
	C_{11} & C_{12}\\
	C_{21} & C_{22})	
	\end{array} \right] 	 ( \begin{array}{c} \beta_1- \hat{\beta_{1}} \\ \beta_2- \hat{\beta_{2}} \end{array}) & = \lambda ( \begin{array}{c} v_1 \\ v_2 \end{array})
\end{align*}
We note that, for the true parameter vector, $\beta_{2} $ is a null vector, by definition. We get the following two equations after some simplification:
\begin{align} \label{eq:kkt_1}
	C_{11}(\beta_{1} - \hat{\beta_1}) - C_{12} \hat{\beta_2} &= \lambda v_1 \\
	C_{21}(\beta_{1} - \hat{\beta_1}) - C_{22} \hat{\beta_2} &= \lambda v_2
\end{align}
After simplification of the first equation we get
\begin{align*}
	\hat{\beta_1}-\beta_{1} = C^{+}_{11} (  C_{12} \hat{\beta_2} +\lambda v_1)\\
	\end{align*}
	Substituting value of $\hat{\beta_1}-\beta_{1}$, in the second equation 
	\begin{align*}
	C_{21} C^{+}_{11} (  C_{12} \hat{ \beta_2 } +\lambda v_1) - \Sigma_{22} \beta_2 &= \lambda v_2 
	\end{align*}
	by multiplying both the sides with $\hat{\beta}_2^T$
	\begin{align*}
	\hat{\beta}_2^T (C_{22} - C_{21} C^{+}_{11} C_{12} ) \hat{ \beta_2 } &= -\lambda \| \beta_2 \|_1 - \hat{ \beta_2 }^T C_{21} C^{+}_{11} \lambda v_1
	\end{align*}
Applying holders inequality for the term $\hat{ \beta_2 }^T C_{21} C^{+}_{11}\lambda v_1$ on RHS we get
\begin{align*}
	\hat{ \beta_2 }^T C_{21} C^{+}_{11} \lambda v_1 & \leq \lambda \| \hat{ \beta_2 } \|_1 C_{21} C^{+}_{11} v_1 \|_{\infty} \\
	\implies & \leq  \lambda \| \hat{ \beta_2 } \|_1.
\end{align*}
We get the following expression after substitution,
\begin{align*}
	\hat{\beta}_2^T (C_{22} - C_{21} C^{+}_{11} C_{12} ) \hat{ \beta_2 } \leq - \lambda \| \hat{ \beta_2 } \|_1.
\end{align*}
	
Since $\lambda \hat{ \beta_2 } \|_1 > 0 $ we get the following inequality, 
\begin{align*}
	\hat{\beta}_2^T (C_{22} - C_{21} C^{+}_{11} C_{12} ) \hat{ \beta_2 } \leq 0
\end{align*}
The matrix $(C_{22} - C_{21} C^{+}_{11} C_{12} )$ is a positive semi-definite, we have arrived at a contradiction. Therefore $\hat{\beta}_{S^c} = 0$, for any Lasso solution it is true. Hence the Lasso correctly identifies all the zero components. 
Hence giving the similar argument as lemma (\ref{lemma:lasso_dual}), it can be shown that $|X_j^T \hat{\theta}| < \lambda$ for all $j \in S^c$. Therefore PIC implies dual lasso selects the true active set.

Now, we assume that dual lasso selects the true active set, and we will show that the PIC holds. 
It is given that $|X_j^T \hat{\theta}| < \lambda$ for all $j \in S^c$. Therefore for any beta solution  $\hat{\beta}_{S^c} = 0$. Using the KKT condition again, and substituting $\hat{\beta}_2 = 0$ in (\ref{eq:kkt_1}), we get the following expression.
\begin{align*} 
	C_{11}(\beta_{1} - \hat{\beta_1}) &= \lambda v_1 \\
	C_{21}(\beta_{1} - \hat{\beta_1}) &= \lambda v_2
\end{align*}
After solving the above we have
\begin{align*} 
	C_{21} C_{11}^+ \lambda v_1 &= \lambda v_2
\end{align*}
Since  $|X_j^T \hat{\theta}| < \lambda \; for j \in S^c$ , therefore $\| v2 \|_{\infty} < 1$ and we have the following PIC.
\begin{align*} 
	\| C_{21} C_{11}^+ \lambda sign(\beta_1) \|_{\infty} & <  \lambda \\
	\implies \| C_{21} C_{11}^+ sign(\beta_1) \|_{\infty} <  1
\end{align*}
\end{proof}


\bibliography{niharika}
\end{document}